\documentstyle[12pt]{article}
\begin{document}

\title{Radiative Corrections to the Hadronic Cross--Section \\
Measurement at DA$\Phi$NE }

 \author{V.A. Khoze $^{a)}$, M.I. Konchatnij $^{b)}$, N.P.
 Merenkov $^{b)}$, G. Pancheri $^{c)}$, \\ L. Trentadue $^{d)}$ and
 O.N. Shekhovzova $^{b)}$}
 \date{}
 \maketitle
 \begin{center}

{\small {\it $^{a)}$ Department of Physics, University of Durham,
\\ Durham, DH1
 3LE, England } \\}
{\small {\it $^{b)}$ National Science Centre "Kharkov Institute of
Physics and
  Technology,\\ 61108 Akademicheskaya 1, Kharkov, Ukrainia} \\ }
{\small {\it $^{c)}$ INFN Laboratori Nazionali di Frascati,\\ P.O. Box 13,
I00044 Frascati, Italy } \\ } {\small {\it ${^d)}$ Dipartimento di Fisica,
Universit\'a di Parma and INFN, \\ Gruppo Collegato di Parma, 43100 Parma,
Italy } \\ } \end{center}

\vspace{0.5cm}
\begin{abstract}
The hadronic invariant mass distribution for the process of
electron--positron annihilation into a pair of charged pions accompanied by
a photon radiated from the initial state has been studied for the region of
DA$\Phi$NE energies. The Born cross--section and the electromagnetic
radiative corrections to it are calculated for realistic conditions of the
KLOE detector. The dependence on the physical parameters which define
the  event selection is obtained.
\end{abstract}

 \section{Introduction}

\hspace{0.7cm}

 The idea to use  radiative events in electron--proton and
electron--positron interactions to expand the experimental possibilities
for studies of different topics in high energy physics has become quite
attractive in the last years. Different aspects of utilizing the
radiative photons are now under intensive discussion.

Radiative events have been used already to measure the structure function
$F_2(x,Q^2)$ at HERA [1]. The corresponding experimental setup takes
advantage of a circle--shaped photon detector (PD) in a very forward
direction, as seen from the incoming electron beam. The PD measures the photon
energy of all photons hitting it. The Born cross--section for such
experimental conditions has been computed in [2], and  further
theoretical study has been performed to calculate the radiative
corrections (RC) to the Born cross--section [3].

The possibility to undertake the $\Upsilon$--spectroscopy studies at the
$\Upsilon(4S)$ energy using the emission of a hard photon from the electron or
positron has been considered in [4]. Estimates performed in this paper
have demonstrated the feasibility of using the radiative photon events for
the investigation of  bottomonium spectroscopy at $B$--factories.

Photon radiation from the initial $e^+e^-$--state in the events with
missing energy has been successfully used at LEP for the measurement of
the number of light neutrinos and for searches of  new physics signals,
see Ref. [5].

Recently,  proposals
to scan the hadronic cross--section
$\sigma_h=\sigma(e^+e^-\rightarrow hadrons)$ at DA$\Phi$NE
energies through one  such radiative process [6,7] have been put forward.
The strong motivation for such proposals lies in  the fact that
the measurement of $\sigma_h$, if performed below the one percent
accuracy level, would allow an instructive  test of the Standard Model via
a precise determination of the anomalous magnetic moment $a_{\mu}=
(g-2)_{\mu}/2$
and the running electromagnetic coupling  at the $Z$--peak $\alpha(M_Z^2)$
[7,8].  On-going experiments in Brookhaven will soon
reduce the experimental error on $a_{\mu}$ below the precision with which
the electroweak contribution to this quantity is known, and could in principle
make tests of new physics. Unfortunately, this cannot yet be envisaged, since
the theoretical error on
$a_{\mu}$ comes mainly from the
uncertainty of the hadronic vacuum polarization contribution in the energy
region below and
around 1 GeV, where the hadronic contribution  to the photon self
energy cannot be calculated unambiguously
within the framework of perturbative $QCD$.  Instead, this
contribution is obtained via
dispersion relations  for the cross-section $\sigma(e^+e^- \rightarrow
hadrons)$ [8,9].
A precise experimental
determination of this quantity appears therefore the only means, at present,
to
reduce the theoretical error on $a_{\mu}$. There are two possible
ways to measure unambiguously this
cross-section in the energy
region of interest, the direct scanning, as presently done at Novosibirsk
[10], and the radiative return method. The radiative return method has
a much smaller cross-section and,
in order to have a statistical error in
the necessary range, i.e. a fraction of a percent,
 it requires much more machine luminosity
than
the direct scanning, which is in principle
the easiest to perform.
Unfortunately, the reduction of the error through
such direct measurement is not to be attained soon. Indeed,
the
precision attainable at VEPP2-M is limited by the machine luminosity, while
DA$\Phi$NE, which has a much higher design luminosity, is planning to
operate at the c.m. energy $\sqrt{s}=M_{\Phi}$ for the next few years.
However, one can still make use of the planned DA$\Phi$NE facility for this
measurement in the near future, through
the radiative return method, recently proposed as mentioned [6,7].
On the theoretical side,
in order to reduce the
systematic errors, it is necessary to perform radiative correction to at
least the percent level to the process
\begin{equation}\label{1}
 e^-(p_1) + e^+(p_2) \rightarrow \gamma(k) + hadrons(q) \ .
 \end{equation}
When
DA$\Phi$NE operates at the $\Phi$--peak, the hadronic final state is dominated
by the $\rho$-resonance decay products and the  proposals
to perform the experimental scanning of the
$\pi^+\pi^-\gamma$
final state [6]  to contribute to the reduction of the error on
$a_{\mu}$ has made the detailed analysis of RC to process (1) a subject of
theoretical
efforts.

The Born cross--sections for the radiative process of
electron--positron annihilation into a pair of charged fermions or
scalar bosons were first calculated in [11]. This topics was subsequently
considered in several papers, see, for example, Refs. [12,13].

In Ref.[14] the RC to total hadronic cross--section of process (1)
with ISR were calculated analytically for the case when the PD
measures the energies of all photons emitted in the narrow cone
along the direction of the electron beam. These corrections include
the first--order contribution with the next--to--leading accuracy
and the high--order terms computing within the leading accuracy.
At present such kind of PD is not the case for
DA$\Phi$NE. The KLOE detector allows to tag photons only outside
a blind zone defined by two narrow cones along both, electron and
positron beam directions. In addition, events with two hard
photons tagged by the PD are rejected. Therefore, generally
speaking the RC depend on "soft" (because of the photon energy
selection in the PD) and "collinear" (because of the PD geometry)
radiation parameters. Note that, as discussed in Section 3,
the Born cross--section depends
on collinear parameters only.

An analytical calculation of the first--order RC to the
distribution over the tagged photon energy for the KLOE--type
detector has been performed in [15], and an analysis of the
$\pi^+\pi^- \gamma$ final state has been carried out in [16] using
the Monte Carlo event generator for the evaluation of the RC given
in [17].

The calculations of the RC performed in [15] do not take into
account some specific (but essential) details of the
event selection in the proposed experiment with the KLOE detector [6].
In this paper we calculate the distribution over the hadronic invariant
mass  in process (1) in the Born approximation
and compute analytically the RC to this distribution
accounting for the cuts discussed in Section 2.
\section{Event selection in KLOE}

\hspace{0.7cm}

The KLOE detector allows to measure independently the energy of the photon
$\omega$ with the  calorimeters (QCAL and EMCAL) and the invariant
mass of the charged pions $q^2$ with the drift chamber (DC). The strategy
of the  experiment will be based on the measurements of the $q^2$ of the
two pions with the DC which indirectly allows to reconstruct $\omega$. The
much higher accuracy of the DC measurements as compared to the finite
resolution of the electromagnetic calorimeter (EMCAL) is
the basis for such a strategy. An attractive advantage of this approach
is that, in principle, it does not require
corrections of the measured distributions for the effects, caused by the
experimental resolution of the calorimeter, the so--called deconvolution
procedure, see Ref. [7].

Let us define the total 4--momentum of the initial electron and positron as
$$p_1+p_2 = (2E,\vec P_{\Phi}) \ , $$
where $E$ is the beam energy, $\vec P_{\Phi}$ is the momentum of the
${\Phi}$ and $|\vec P_{\Phi}|$ = 12.5 MeV in the $X$--direction [6]. Note
that in the laboratory frame the Lorentz boost of the ${\Phi}$ is
accounted for. In the interaction point the electron and positron
exercise not exactly a head-on collision but there is a small
beam crossing angle of
order $|\vec P_{\Phi}|/2E$ relative to the $Z$ axis and
\begin{equation}\label{4-momentum} p_1=\bigl(E,\frac{|\vec P_{\Phi}|}{2},
0 , P_z\bigr)\ , \ \ p_2=\bigl(E,\frac{|\vec P_{\Phi}|}{2}, 0 ,-
  P_z\bigr)\ , \ \ P_z = E\bigl(1-\frac{|\vec P_{\Phi}|^2}{8E^2}\bigr) .
  \end{equation}
Here we define the $XZ$ as the $(\vec p_1, \vec p_2)$ -- plane, and $Z$
as the symmetry axis of the PD.

In spite of its smallness, the quantity $|\vec P_{\Phi}|/E$ should be
taken into account in a high precision determination of the photon energy
and in the calculation of the cross--section of process (1).

In the single photon emission events the photon energy $\omega$ can
be reconstructed directly  from
$ q^2 = (p_1+p_2-k)^2$
\begin{equation}\label{photon energy}
 \omega = \frac{4E^2 -|\vec P_{\Phi}|^2-q^2}
{2(2E-|\vec P_{\Phi}|\sin{\theta}\cos{\varphi})}\ ,
\end{equation}
where $\theta (\varphi)$ is polar (azimuthal) angle of a photon in
the laboratory frame. We see that because of the difference
between the laboratory and the centre-of-mass frames one can
reconstruct the photon energy in (1) only if the exact angular
orientation of the emitted photon is known.

A systematic error could arise due to the events with the multiple photon
emission. In order to reject these events as well as to decrease the
background caused by the final--state radiation the following event
selection cuts are imposed [6]
\begin{equation}\label{4,restriction on the
event selection} \Omega -|\vec K| \leq \eta \ , \ \omega\geq\omega_m \ , \
\omega_m = 50 MeV\ , \ \ \eta = 10 MeV \ , \end{equation} where $\Omega$
($\vec K$) is the energy (3--momentum) of all emitted photons, assuming
that only one hard photon with the energy $\omega$ is tagged by the PD.
Here $\omega_m$ denotes  threshold energy for this photon. These
restrictions are based on the predictions of the Monte Carlo events
generator described in Ref. [16].

The first inequality in (4) is the reduced form of the constraint \\

$\ \ \ \ \ \ \ \ \ \ \ \ \ \ \ \ \ \ \ \ \ \ \ \ \ \ \ M_{\Phi} - E_+ -
E_- - |\vec P_{\Phi} - \vec p_+ - \vec p_-| < \eta \ , $ $\ \ \ \ \ \ \ \
\ \ \ \ \ \ \ \ \ \ \ \ \ \ \ \ \ \ \ \ \ \ \ \ \ \ $ (4a)
where $M_{\Phi}$ is
the mass of $\Phi$--meson and $E_{+,-} \ (\vec p_{+,-})$ is the energy
(3--momentum) of $\pi^+,\ \pi^-$ for $\pi^+\,\pi^- + n\gamma$ events with
$n \geq 1.$

Photon tagging by the QCAL
calorimeter, which surrounds the blind zone and covers the angles from
$\theta_m$ up to $20^o$ with respect to the electron beam direction as
well as the symmetrical angles along the positron beam, can be done for
the photon energy above the threshold $\omega_{min}^{qc} = 1 MeV$ (here
$\theta_m$ is the aperture of the blind zone ). The corresponding
threshold for the EMCAL calorimeter, which covers the angles from $20^o$
up to $40^o$ with respect to both the electron and the positron beam
directions, is $\omega_{min}^{ec} = 5 MeV)$ [6].  As we noted above, the
events with the two hard photons inside the PD are assumed to be rejected.
Therefore, when calculating the RC, one has to take into account the
possibility that a soft photon with the energy $\omega_1$ (additional to
the tagged one) hits the detector, but is not registered. Thus, we can
write the following constraints on the energy $\omega_1$ and the
radiation angle $\theta_1$ of an additional soft photon inside the
detector \begin{equation}\label{5,event selection} \omega_1 < \Delta_1E \
, \ if \ \ 160^{\circ} < \theta_1 < \pi - \theta_m \ \ and \ \ \theta_m <
\theta_1 < 20^{\circ}\ , \ \ \Delta_1 = \frac{\omega_{min}^{qc}}{E} \simeq
0.2\cdot 10^{-2};\ \end{equation} $$ \omega_1 < \Delta_2E \ , \ if \ \
40^{\circ} > \theta_1 > 20^{\circ} \ , \ and \ \ 160^{\circ} > \theta_1 >
140^{\circ} \ , \ \ \Delta_2= \frac{\omega_{min}^{ec}}{E}\simeq 10^{-2}\ ,
$$ where $\theta_m$ is for about $10^{\circ}$ and $\theta_1$ is defined
relative the Z axis.

In the following Section, we shall compute the distribution over the
hadronic invariant
mass in process (1) in the Born approximation.

\section{Born approximation}

\hspace{0.7cm}

To lowest order in $\alpha$, the differential
cross--section for process (1) with respect to the tagged hard
photon has been calculated in [11], and here we reproduce the
expression for an arbitrary hadronic final state.

The general formula for the differential cross--section in the Born
approximation can be written as
\begin{equation}\label{6}
d\sigma^B = \frac{2\pi^2\alpha^2}{S}|M|^2\frac{\alpha}{4\pi^2}\frac{d^3k}
{\omega}d\Gamma \ ,
\end{equation}
where $\alpha$ is the electromagnetic coupling , $ S=2(p_1p_2) $ and
$$d\Gamma = (2\pi)^4\delta(q-\sum
q_f)\prod\frac{d^3q_f}{2\varepsilon_f(2\pi)^3} $$
is the phase space factor for the final hadrons, $q_f$ is the 4--momentum of an
individual  hadron. The squared matrix element on the right--hand side of
Eq.(6) can be written in terms of the electronic
and hadronic tensors $L_{\mu\nu}^{^{\gamma}}$ and $H_{\mu\nu}$ as
\begin{equation}\label{7}
|M|^2 = \frac{4}{q^4}L_{\mu\nu}^{^{\gamma}}H_{\mu\nu} \ .
\end{equation}
The subscript $\gamma$ in the electronic current tensor indicates that
here we are dealing with ISR events in process (1).

The differential cross section for radiative events can be obtained by
integrating  over all hadronic final
states. This can be performed by using the well known relation
\begin{equation}\label{8}
\sum_h\int H_{\mu\nu}(q)d\Gamma = F_h(q^2)\widetilde g_{\mu\nu} \ , \ \
\widetilde g_{\mu\nu} = g_{\mu\nu} - \frac{q_{\mu}q_{\nu}}{q^2} \ ,
\end{equation}
where the function $F_h(q^2)$ carries all the information about
 the non--radiative
hadronic cross--section $\sigma_h(q^2)$. For the case of annihilation into
a charged pion pair
\begin{equation}\label{9}
F_h(q^2) =
\frac{q^2|F_{\pi}(q^2)|^2}{24\pi}\bigl(1-\frac{4\mu^2}{q^2}\bigr)^{^{3/2}}
\ , \end{equation} where $\mu$ is the pion mass and $F_{\pi}(q^2)$ is the
pion electromagnetic form factor.

The leptonic tensor can be presented as [11,18]
\begin{equation}\label{10}
L_{\mu\nu}^{^{\gamma}} = \frac{(S+T_1)^2+(S+T_2)^2}{T_1T_2}\widetilde
g_{\mu\nu} + \frac{4q^2}{T_1T_2}(\tilde p_{1\mu}\tilde p_{1\nu}+
\tilde p_{2\mu}\tilde p_{2\nu}) \ , \ \ \tilde p = p-\frac{pq}{q^2}q \ ,
\end{equation}
where we introduced the following notations
$$T_1 = -2p_1k\ = -\omega\bigl(2E-2P_z\cos{\theta}-|\vec
P_{\Phi}|\sin{\theta}\cos{\varphi}\bigr) \ ,$$
$$T_2 = -2p_2k\ = -\omega\bigl(2E+2P_z\cos{\theta}-|\vec
P_{\Phi}|\sin{\theta}\cos{\varphi}\bigr) \ ,$$
$$S=2p_1p_2 \ = 4E^2-|\vec P_{\Phi}|^2 \ , \ \ q^2=S+T_1+T_2 \ .$$
In the expression for the leptonic tensor we have
neglected terms of relative order $m^2/|T_1|$ and $m^2/|T_2|$ (here $m$ is
the electron mass) because for the KLOE detector these cannot exceed
$m^2/(E^2\theta_m^2) \simeq 10^{-4}$.

Taking into account that $$L_{\mu\nu}^{^{\gamma}}\widetilde
g_{\mu\nu} = 2\frac{(S+T_1)^2+ (S+T_2)^2}{T_1T_2}$$ we can present
the Born cross--section as (see also Ref. [11])
\begin{equation}\label{11}
d\sigma^B = \sigma(q^2)\frac{\alpha}{2\pi^2}\frac{(S+T_1)^2+(S+T_2)^2}
{T_1T_2}\frac{d^3k}{S\omega} \ , \ \ \sigma(q^2) =
\frac{\pi\alpha^2|F_{\pi}(q^2)|^2}{3q^2}
\bigl(1-\frac{4\mu^2}{q^2}\bigr)^{^{\frac{3}{2}}}\ .
\end{equation}

Let us multiply the right--hand side of Eq.(11) by
$$dq^2\delta(4E(E-\omega)-|\vec P_{\Phi}|^2-q^2+2\omega|\vec P_{\Phi}|
\sin{\theta}\cos{\varphi}) $$
and use the $\delta$--function to perform the integration over $d\omega.$
Imposing the threshold restriction (4) for the events with the emission of
a single photon
$$\omega\geq\omega_m$$ we arrive at
\begin{equation}\label{12}
\frac{d\sigma^{^B}}{dq^2} = \frac{\alpha}{2\pi^2}\sigma(q^2)\frac{(S-q^2)
d\cos{\theta}d\varphi}{4S(2E-|\vec
P_{\Phi}|\sin{\theta}\cos{\varphi})^2}\frac{(S+T_1)^2+(S+T_2)^2}{T_1T_2}
\end{equation}
$$\Theta\bigl(\frac{S-q^2}{2(2E-|\vec P_{\Phi}|\sin{\theta}\cos{\varphi})}-
\omega_m\bigl)\ .$$

In principle one can perform the angular integration on the right--hand side of
Eq.(12) numerically. The analytical integration is complicated because of the
$\Theta$--function. This results in nontrivial limits for the angular
integration. To derive them let us first examine the quantity
$$ D= \frac{4E(E-\omega_m)-q^2-|\vec P_{\Phi}|^2}{2\omega_m|\vec
P_{\Phi}|} \ . $$

If $D>1$ then the emission of a photon is allowed in all available angular
phase space. For $$ 1  >  D  >  \sin{\theta_m}$$ two options appear,
namely \begin{equation}\label{13,a} 2\pi > \varphi > 0 \ , \ \
\arcsin{D} > \theta > \theta_m\ , \ \
\pi-\theta_m > \theta > \pi-\arcsin{D}
\end{equation}
and
\begin{equation}\label{14,b}
\arccos{\frac{-D}{\sin{\theta}}}  > \varphi  > 0\ , \ \ 2\pi  >
\varphi  > 2\pi-\arccos{\frac{-D}{\sin{\theta}}} \ , \ \
\pi-\arcsin{D} > \theta > \arcsin{D} \ .
\end{equation}

When $\sin{\theta_m} > D > -\sin{\theta_m}$ or $-\sin{\theta_m} > D > -1$
the limits for the azimuthal integration are the same as in Eq.(14)
but the polar angles are different
\begin{equation}\label{15}
\sin{\theta_m} > D > -\sin{\theta_m}\ , \ \ \pi-\theta_m > \theta > \theta_m \ ,
\end{equation}
whereas at
\begin{equation}\label{16}
-\sin{\theta_m} >  D  >  -1 \ , \ \ \pi-\arcsin{(-D)} > \theta > \arcsin{(-D)}
\ .  \end{equation}

Single photon emission with energy $\omega > \omega_m$ is not
allowed if $$D < -1 \ .$$

Using now the angular constraints  given by Eqs.(13)--(16)
one can perform the angular integration in Eq.(12) analytically (at least
over the azimuthal angle). The result can be presented in the following form
\begin{equation}\label{17,full Born}
\frac{d\sigma^{^{B}}}{dq^2} = \frac{d\sigma^{^{B}}(D > 1)}{dq^2} +
\frac{d\sigma^{^{B}}_a(1 > D > s_m)}{dq^2} + \frac{d\sigma^{^{B}}_r}{dq^2} \ .
\end{equation}

The quantity $d\sigma^{^{B}}(D > 1)/dq^2$ corresponds to events with $D > 1$
when all radiation angles for the tagged photon are allowed. It reads
\begin{equation}\label{18, integration for D > 1}
\frac{d\sigma^{^B}(D > 1)}{d\ q^2}=
\frac{\alpha}{2\pi}\frac{\sigma(q^2)}{2E^2}
\Bigl\{\bigl(\frac{S}{S-q^2}-1+\frac{S-q^2}{2S}\bigr)\Bigl[2\ln\frac{1+c_m}
{1-c_m}+\frac{|\vec
P_{\Phi}|^2}{2E^2}\bigl(\ln\frac{1+c_m}{1-c_m}+\frac{c_m}{s_m^2}\bigr)\Bigr]
\end{equation}
$$-\frac{S-q^2}{S}c_m\bigr[1+\frac{|\vec
P_{\Phi}|^2}{8E^2}(3-c_m^2)\bigr] \Bigr\} \ , $$ where
$c_m=\cos{\theta_m} \ , \ s_m = \sin{\theta_m}.$ The quantity
$d\sigma^{^{B}}_a(1 > D > s_m)/dq^2$ describes events with
$1 > D > s_m$ for which  full coverage in the azimuthal angle is
allowed. It has a structure which is very close to the right--hand
side of Eq.(18) and reads
\begin{equation}\label{19, integration for1 >  D > s_m(a)}
\frac{d\sigma^{^B}_a(1 > D > s_m)}{d\ q^2}=
\frac{\alpha}{2\pi}\frac{\sigma(q^2)}{2E^2}
\Bigl\{\bigl(\frac{S}{S-q^2}-1+\frac{S-q^2}{2S}\bigr)\Bigl[2\ln\frac{(1+c_m)
(1-c_d)}{(1-c_m)(1+c_d)}+
\end{equation}
$$\frac{|\vec
P_{\Phi}|^2}{2E^2}\bigl(\ln\frac{(1+c_m)(1-c_d)}{(1-c_m)(1+c_d)}+\frac{c_m}
{s_m^2} - \frac{c_d}{s_d^2}\bigr)\Bigr]
-\frac{S-q^2}{S}(c_m-c_d)\bigr[1+\frac{|\vec
P_{\Phi}|^2}{8E^2}(3-c_m^2-c_d^2-c_mc_d)\bigr] \Bigr\} \ , $$
where $s_d = D\ , \ c_d = \sqrt{1-D^2}.$

The contribution of the remaining regions (see Eqs.(14), (15) and (16)) is
described by the quantity $d\sigma^{^{B}}_r/dq^2.$ We integrate it only
over the azimuthal angle
 and arrive at
\begin{equation}\label{20, rest Born contribution}
\frac{d\sigma^{^{B}}_r}{dq^2} =
\frac{\alpha}{\pi}\sigma(q^2)d\cos{\theta}\Bigl[
\Bigl(\frac{S}{S-q^2}-1+\frac{S-q^2}{2S}\Bigr)\frac{2}{\pi
P_z\cos{\theta}}(\Phi_- -\Phi_+) -
\end{equation}

$$\frac{S-q^2}{\pi S(4E^2-|\vec
P_{\Phi}|^2\sin^2{\theta})}\Bigl(\frac{|\vec
P_{\Phi}|\sqrt{\sin^2{\theta}-s_d^2}}{2E-|\vec P_{\Phi}|s_d} +
4E\Phi\Bigr)\Bigr] \Theta_r\ ,$$ where $$\Phi_{\pm} =
\frac{1}{\sqrt{(2E\pm2P_z\cos{\theta})^2-|\vec
P_{\Phi}|^2\sin^2{\theta}}} \arctan{\sqrt{\frac{(2E\pm
2P_z\cos{\theta}+|\vec
P_{\Phi}|\sin{\theta})(\sin{\theta}+s_d)}{(2E\pm
2P_z\cos{\theta}-|\vec P_{\Phi}|\sin{\theta})(\sin{\theta}-s_d)}}}
\ . $$ Note that on the right--hand side of Eq.(20)  $\sin{\theta}$
always exceeds $s_d.$ The quantity $\Phi$ can be obtained from
$\Phi_+$ (or $\Phi_-$) by $$\Phi = \Phi_{\pm}(P_z=0) \ . $$ The
function $\Theta_r$ on the right--hand side of Eq.(20) is
introduced to define the upper limits of the variable
$\cos{\theta}$ at different values of  $D.$ It can be written
explicitly as
\begin{equation}\label{21} \Theta_r =
\theta(c_d-\cos{\theta})[\theta(1-D)\theta(D-s_m) +
\theta(-s_m-D)\theta(D+1)] +
\end{equation}
$$\theta(c_m-\cos{\theta})\theta(s_m-D)\theta(s_m+D) $$
provided that the minimal value of $\cos{\theta}$ equals to zero.
One can verify that in the limit $|\vec P_{\Phi}|=0$ when only the
region $D > 1$ contributes (all emission angles are allowed) the
right--hand side of Eq.(17) coincides with the well known
expression, see, for example, Refs. [15,16]
\begin{equation}\label{22}
\frac{d\sigma^B}{dx}=\frac{\alpha}{2\pi}\sigma(q^2)2\bigl[\frac{1+(1-x)^2}{x}
\ln\frac{1+c_m}{1-c_m}-xc_m\bigr] \ ,
\end{equation}
where in this limit $x=\omega/E = (4E^2-q^2)/(4E^2)$ .

Note that an account for the $|\vec P_{\Phi}|/E$ effects is mainly
essential for the reconstruction of the tagged photon energy (see Eq.(3))
if one wishes to guarantee the one percent accuracy level.

\section{Radiative corrections}

\hspace{0.7cm}

The proposed high accuracy measurement of the pion contribution to the
hadronic cross--section at DA$\Phi$NE [6] by using radiative
events in  process (1), requires an adequately high
precision of the theoretical predictions.
These  have to take into account at least the first--order QED
radiative corrections. The first--order RC to $d\sigma^B/dq^2$ include
the real and virtual soft photon
contributions in the overall phase
space as well as the hard real contribution from the region where the PD
does not tag a photon. Since the effect caused by the deviation of the
laboratory frame from the centre-of-mass frame
is small ( of relative order $|\vec P_{\Phi}|/E$) it may be neglected in
the calculation of the RC. Within this approximation we define in
this section the invariants
$$ s = S(|\vec P_{\Phi}|=0)\ , \ \ t_{1,2} = T_{1,2}(|\vec P_{\Phi}|=0)
\ .$$

\subsection{Virtual and soft corrections}

\hspace{0.7cm}

The RC due to  virtual photon emission can be computed
employing the results of Ref. [18] (see also Ref. [19]) where the
one--loop corrected Compton tensor with a heavy photon has been
calculated for the scattering channel. In oder to obtain the
corresponding results for the annihilation channel it is
sufficient to make the substitutions $$p_2\rightarrow-p_2 \ , \ \
u\rightarrow s \ , \ \ s\rightarrow t_2 \ , \ \ t\rightarrow t_1
$$ in all formulae of Ref. [18].

In accordance with Ref. [18] the contribution to the differential
cross--section for process (1) due to virtual and soft
photon emission can be written as \begin{equation}\label{23}
d\sigma^{^{V+S}} = \frac{\alpha^2}{8\pi^3}\sigma(q^2)\bigl[\rho
L_{\mu\nu}^{^{\gamma}}+T_{\mu\nu}\bigr]\widetilde
g_{\mu\nu}\Theta\bigl(\frac{s-q^2}{4E}-\omega_m\bigr)\frac{d^3k}{s\omega}
\ , \end{equation} where the quantity $\rho$ absorbs all infrared
singularities. It can be presented as a sum of two contributions
$$\rho = \rho_V +\rho_S \ ,$$ where $\rho_V$ arises due to one--loop
virtual corrections and $\rho_S$ -- due to  soft photon contributions.
For the quantity $\rho_V$ we can use an expression derived in [18]
\begin{equation}\label{24, infrared loop-correction}
\rho_V = 4\ln\frac{\lambda}{m}(L_s-1)-L_s^2 +3L_q+\frac{4\pi^2}{3}
-\frac{9}{2} \ , \ \ L_s=\ln\frac{s}{m^2}\ , \ L_q=\ln\frac{q^2}{m^2}\ .
\end{equation}
Concerning the quantity $\rho_S$, the results of  Ref. [18]
are not valid , since in our case the experimental requirements for
the softness of an additional photon inside the PD depend on its
polar angle, see Eq.(5). Note that the parameters $\Delta_1$ and
$\Delta_2$ in (5) are the physical ones and they will appear
explicitly into the final expression for the RC. If an additional
soft photon is outside the PD we can use an arbitrary small
parameter $\Delta$ to define its maximum energy fraction. This
parameter is an auxiliary one, and it disappears in the final
result for the RC due to the possibility of an additional untagged
hard photon emission outside the PD (see below).

When evaluating the soft photon corrections we present the corresponding
cross--section in the factorized form
\begin{equation}\label{25}
d\sigma^S = d\sigma^B\delta\ , \ \ \delta =
-\frac{\alpha}{4\pi^2}\int\frac{d^3k_1}{\omega_1}\Bigl(\frac{p_1}{p_1k_1}-
\frac{p_2}{p_2k_1}\Bigr)^2 \ , \ \ \omega_1 = \sqrt{(\vec
k_1^2+\lambda^2)} \ ,
\end{equation}
where $k_1$ is the 4--momentum of an additional soft photon. Such
factorized form is valid if $\sigma(q^2)$ is a
smooth function of $q^2$ , see Refs. [15,20]. In the case under
consideration, the width of the $\rho$--resonance is large enough, and
approximation (25) is justified.

One can verify, whether restriction (4) $(\Omega -|\vec K|
 < \eta)$ affects this form. Obviously, Eq.(25) is
valid in the case of a very soft additional radiated photon.
It is, therefore, sufficient to examine its impact for the
maximum allowed energy of an additional soft photon. According to Eq.(5)
this maximum energy is $\Delta_2E$. The above--mentioned restriction can
be presented as $$(\omega +\omega_1 -\eta)^2  <
\omega^2+\omega_1^2+2\omega\omega_1\bar{c} \ , $$ where $\bar{c}$
is the cosine of the angle between the vectors $\vec k$ and $\vec
k_1.$ Since for the
collinear photons $\gamma(k)$ and $\gamma(k_1)$ the constraint
(4) is always fulfilled, one can check its validity for the maximal
angle or for $\bar{c}=-1.$ Setting $\bar{c}=-1$ and
$\omega_1=\Delta_2E$ in the previous equation we obtain
\begin{equation}\label{26} (2\omega-\eta)(2\Delta_2E-\eta)  <  0
\ .  \end{equation} Therefore, we have $$\eta > 2\Delta_2E .$$ From
Eqs.(4) and (5) it follows that this restriction is satisfied.
Thus, we can use
representation (25) in all angular phase space for an additional soft
photon.

Integration of Eq.(25) with constraints (5) and

$$ \theta_1 < \theta_m \ , \ \theta_1 > \pi-\theta_m \ , \ \ \omega_1
< \Delta E $$
leads to the following result for the RC due to soft photon emission
\begin{equation}\label{27}
\delta = \frac{\alpha}{2\pi}\rho_S \ ,
\end{equation}
$$ \rho_S = \bigl[4(1-L_s)\ln\frac{\lambda}{\Delta m}+L_s^2-\frac{2\pi^2}{3}
+ 4\bigl(\ln\frac{\Delta_1}{\Delta}\ln\frac{1+c_m}
{1-c_m}+\ln\frac{\Delta_2}{\Delta_1}\ln\frac{1+c_1}{1-c_1}
 + \ln\frac{\Delta}{\Delta_2}\ln\frac{1+c_2}{1-c_2}\bigr)\bigr] \ ,
$$
where $c_1=\cos{20^{\circ}}$ and $c_2=\cos{40^{\circ}}$ for the KLOE
photon detector.  Therefore, for the factor $\rho$, which is the sum of
(24) and (27) we have \begin{equation}\label{28} \rho = 4(L_s-1)\ln\Delta
+ 3L_q +\frac{2\pi^2}{3}-\frac{9}{2}+
4\bigl(\ln\frac{\Delta_1}{\Delta}\ln\frac{1+c_m}
{1-c_m}+\ln\frac{\Delta_2}{\Delta_1}\ln\frac{1+c_1}{1-c_1}
+ \ln\frac{\Delta}{\Delta_2}\ln\frac{1+c_2}{1-c_2}
\bigr).
\end{equation}
Note that quantity $\rho$ coincides with the well--known expression
in the limiting case $\Delta_i = \Delta \ , \ i=1,2 $ (see
e.g. [14,15,18]).

Tensor $T_{\mu\nu}$ on the right--hand side of Eq.(23) has the
structure
\begin{equation}\label{29}
T_{\mu\nu} = T_g\widetilde g_{\mu\nu} + T_{11}\tilde
p_{1\mu}\tilde p_{1\nu}+ T_{22}\tilde p_{2\mu}\tilde p_{2\nu}-T_{12}\tilde
p_{1\mu}\tilde p_{2\nu}-T_{21} \tilde p_{2\mu}\tilde p_{1\nu} \ .
\end{equation} Now after carrying out the contraction of tensors on the
right--hand side of Eq.(27) we arrive at \begin{equation}\label{30}
\frac{d\sigma^{^{V+S}}}{dq^2}  = \frac{\alpha}{2\pi^2}\sigma(q^2)
\frac{(s-q^2)d\cos{\theta}d\varphi}{4s^2} \Theta\bigl(\frac{s-q^2}{4E}-
\omega_m\bigl)
\end{equation}
$$\times\frac{\alpha}{2\pi}
\bigl[\rho\frac{(s+t_1)^2+(s+t_2)^2}{t_1t_2}+T\bigr]\ ,$$
 $$T=\frac{3}{2}T_g-\frac{1}{8q^2}\bigl[T_{11}(s+t_1)^2+T_{22}(s+t_2)^2
+(T_{12}+T_{21})(s(s+t_1+t_2)-t_1t_2)\bigr] \ , \ \ s = 4E^2 \ .$$
As has been already mentioned  for the KLOE detector $|t_{1,2}| >  > m^2$, and therefore
one can neglect all terms proportional to $m^2$ in the expressions for
$T_g$ and $T_{ik}.$ Then  we obtain
\begin{equation}\label{31}
T_g =  -\Bigl[\frac{sq^2}{t_2^2}+\frac{2s(s+t_2)+t_2^2}{t_1t_2}\Bigr]G+
s\bigl(\frac{q^2}{t_1t_2}-\frac{2}{t_1+t_2}\bigr)(L_q-L_s) +\frac{s+t_1}
{t_2}\bigr(\frac{3s}{s+t_2}-1\bigr)(L_q-L_1) +
\end{equation}
$$\frac{s^2-t_2^2}{2t_1t_2}+(t_1\leftrightarrow t_2) \ ,$$
$$T_{11} =
\frac{2}{t_1t_2}\Bigl\{-q^2\bigl(1+\frac{s^2}{t_2^2}\bigr)G-
q^2\bigl(2+\frac{(s+t_2)^2}{t_1^2}\bigr)\widetilde G
+2q^2\bigl[\frac{(s+t_2)^2}{t_1t_2}+\frac{2s}{t_1+t_2}\bigr](L_q-L_s)+$$
$$\frac{4}{t_1+t_2}[s^2-(s+t_2)t_1]\bigl[\frac{q^2}{t_1+t_2}(L_q-L_s)-1\bigr]
+\frac{q^2(s+t_2)^2}{t_1(s+t_1)^2}(2s+3t_1)(L_q-L_2)+ $$
\begin{equation}\label{32}
\frac{q^2}{t_2}(2s-t_2)(L_q-L_1)-4s-2q^2+t_1-\frac{(s+t_2)^2}{s+t_1}\Bigr\}\ ,
\end{equation}

\begin{equation}\label{33}
T_{22} = T_{11}(t_1\leftrightarrow t_2 \ , \ G\leftrightarrow\widetilde
G) \ ,
\end{equation}
$$T_{12}+T_{21} =
\frac{2}{t_1t_2}\Bigl\{\frac{q^2}{t_2^2}(s+t_1)(s-t_2)G +
\frac{q^2}{t_1^2}(sq^2-t_1t_2)\widetilde G
-2q^2\bigl(\frac{sq^2}{t_1t_2}
+\frac{2s-t_2+t_1}{t_1+t_2}\bigr)(L_q-L_s)-$$
$$\frac{4[s^2-(s+t_1)t_2}{t_1+t_2}\bigl[\frac{q^2}{t_1+t_2}(L_q-L_s)-1\bigr]
+\frac{q^2}{(s+t_1)^2}(2s+3t_1)\bigl(t_2-\frac{q^2s}{t_1}\bigr)
(L_q-L_2)- $$
\begin{equation}\label{34}
\frac{q^2(s+t_1)}{t_2(s+t_2)}(2s-t_2)(L_q-L_1)+8s+3t_1-t_2+\frac{2st_2}{s+t_1}
\Bigr\}+(t_1\leftrightarrow t_2) \ ,
\end{equation}
where  the following notation has been introduced
$$G = (L_q-L_s)(L_q+L_s-2L_1) +2\bigl[f(1)+f\bigl(1-\frac{q^2}{s}\bigr)-
f\bigl(1-\frac{t_1}{q^2}\bigr)\bigr], \ \widetilde G =
G(1\leftrightarrow 2) \ , $$
$$ L_1=\ln\frac{-t_1}{m^2}\ , \ \ f(x) =
\int\limits_0^x\frac{dt}{t}\ln(1-t) \ . $$

The Born--like contribution (which is proportional to $\rho$) on the right--hand
side of Eq.(30) absorbs all infrared singularities via the quantities
$\ln\Delta, \ \ln\Delta_1$ and $\ln\Delta_2. $ Concerning the collinear
ones, the Born--like term, being integrated over the angular acceptance of
the KLOE electromagnetic calorimeter, generates a contribution
proportional to $\ln\theta_m$ while the remaining $T$-term -- to both
$\ln\theta_m$ and $(\ln\theta_m)^2.$ This can be easily seen from studying
the asymptotic behaviour of the term given in the second line on the
right--hand side of Eq.(30), for instance, at small values of $|t_1|.$
Neglecting the $|\vec P_{\Phi}|/E$ effects we obtain in this limiting case
\begin{equation}\label{35}
|t_1| = 2\omega E(1-c)\simeq E^2\theta_m^2 \ll \ s\ , |t_2| \ , \ \ t_2 =
-4E^2x \ , \ q^2 = 4E^2(1-x) \ , \ \ c=\cos{\theta}\ . \end{equation} Then
we have $$\frac{\alpha}{2\pi}\Bigr\{2\rho\frac{1+(1-x)^2}{x^2(1-c)} +
\frac{2}{x(1-c)}\Bigl[\frac{1+(1-x)^2}{x}\bigl(\ln(1-x)\ln\frac{x^2(1-c)}
{2(1-x)}-2f(x)\bigr)+\frac{2-x^2}{2x}\Bigr]\Bigr\} \ . $$

\subsection{Radiation of an untagged hard photon outside the PD}

\hspace{0.7cm}

When calculating RC due to the radiation of an additional invisible hard
photon, we have to distinguish between the large $(140^o>\theta_1>40^o)$
and small $(\theta_m > \theta_1$ and $ \theta_1> \pi- \theta_m) $ angle
radiation. For large angles we take into account only the contribution
proportional to $\ln\Delta$ and write it in the form
\begin{equation}\label{36}
\frac{d\sigma^L}{dq^2} = \frac{d\sigma^B}{dq^2}
\frac{\alpha}{2\pi}4\ln\frac{1}{\Delta}\ln\frac{1+c_2}{1-c_2}\ ,
\end{equation}
where the Born cross--section is defined by Eq.(12) with $|\vec P_{\Phi}|
= 0.$

To simplify the calculation of the small--angle contribution we use the
quasireal electron approximation [21]. Of course all the
experimental constraints for the event selection should be taken
into account.

 We begin with the general expression for the cross section describing the
radiation of an untagged hard photon inside the small--angle blind zone

\begin{equation}\label{37,general form for hard photon contribution}
\frac{d\sigma^{^{H}}}{dq^2} =
2\frac{d\sigma^{^{B}}_{sh}}{dq^2}\frac{\alpha
E^2}{4\pi^2}\bigl[\frac{1+(1-z)^2}{(k_2p_1)}-\frac{m^2z(1-z)}{(k_2p_1)^2}
\bigr]dzd\cos{\theta_2}d\varphi_2\Theta_{\eta} \ ,
\end{equation}
where $z$ is the energy fraction of an untagged hard photon $z=\omega_2/E,$
$$\frac{d\sigma^{^{B}}_{sh}}{dq^2} =
\frac{d\sigma^{^{B}}(p_1(1-z),k,p_2)} {dq^2} =
\frac{\alpha}{\pi}\sigma(q^2)\frac{s(1-z)-q^2}{[2E(2-z(1-c))]^2} $$
\begin{equation}\label{38}
\times\frac{(1-z)^2(s+t_1)^2+((1-z)s+t_2)^2}{(1-z)t_1t_2}\frac{d\cos{\theta}}
{(1-z)s}
\Theta\bigl[\frac{s(1-z)-q^2}{2E(2-z(1-c))}-\omega_m\bigr]\ , \ \
c=\cos{\theta}
\end{equation}
is the shifted Born cross--section (with the
substitution $p_1\rightarrow(1-z)p_1$) and $$\Theta_{\eta} = \Theta(|\vec
k + \vec k_2| - (\omega + \omega_2 - \eta))$$ is the reduced form of the
restriction $$\Omega -|\vec K|  <  \eta $$ for the case of one untagged
hard photon with the 4--momentum $k_2 = (\omega_2, \vec k_2).$ Factor 2 on
the right--hand side of Eq.(38) appears because Eq.(37) describes
collinear radiation of an additional hard photon along both electron
and positron directions.

The $\Theta$--function on the right--hand side of Eq.(38) defines
the maximum possible energy fraction $z_{max}$ of an additional hard
untagged photon emitted into the small--angle blind zone. It depends on
the pion invariant mass $q^2$ and the polar angle $\theta$ of the hard
photon hitting PD. Therefore it can be rewritten in the following form
\begin{equation}\label{39}
\Theta\bigl[\frac{s(1-z)-q^2}{2E(2-z(1-c))}-\omega_m\bigr] =
\Theta(z_{max}-z) \ , \ \
z_{max} =
\frac{s-q^2-4E\omega_m}{s-2E\omega_m(1-c)}\ .
\end{equation}

The $\Theta_{\eta}$--function on the right--hand side of Eq.(37)
leads to the nontrivial correlations between the limits for
variables $z, \theta_2$ and $\varphi_2.$ (Here we use the
coordinate frame where $Z$--axis is the electron  direction and
define the $(\vec p_1, \vec k)$ plane as $XZ$ ). These limits can
be understood in following way.  First one needs to analyze the
quantity $$B=\frac{z(1-c_2)E-\eta}{zs_2E} \ ,$$ where
$s_2=\sin{\theta}\sin{\theta_2} , \ c_2 =
\cos{\theta}\cos{\theta_2}\ .$ If  $$B > 1$$ then all azimuthal angles
for an untagged  hard photon are allowed, and we face two options
for its polar angle and the energy fraction
\begin{equation}\label{40} I_a-region : \ \ 0 < \theta_2 < \theta_m \
, \ \ \Delta <  z  <  \bigl[z_{max}, \frac{\eta}{E(1-c_+)}\bigr] \ ,
\end{equation}
\begin{equation}\label{41, Ib-region}
I_b-region: \ \ 0 < \theta_2 <
\arccos{\bigl(1-\frac{\eta}{zE}\bigr)}-\theta \ , \ \frac{\eta}{E(1-c_+)}
<  z  <  \bigr[z_{max}, \frac{\eta}{E(1-c)}\bigr] \ .  \end{equation}
where $c_+ = \cos(\theta +\theta_m) $ and $[a,b]$ is min($a$,$b).$

If the value of the quantity
$B$ corresponds to $$1 >  B  >  -1 \ ,$$ then not all azimuthal angles
for an untagged photon are allowed. In this case we obtain the
following constraint \begin{equation}\label{42} 0  <  \varphi_2  <
\arccos{B} \ , \ \ 2\pi  >  \varphi_2  >  2\pi - \arccos{B} \ .
\end{equation} The region defined by Eq.(42) is symmetric relative to the
plane $(Z,X),$ which contains the momentum of  the photon hitting the PD .
For this case there are also two possibilities for the limits for
$\theta_2$ and $z$
\begin{equation}\label{43, II - region for angles and
z} II_a -region: 0 < \arccos{\bigl(1-\frac{\eta}{zE}\bigr)} - \theta  <
\theta_2  < \theta_m \ , \ \ \frac{\eta}{E(1-c_+)}  <  z  <
\bigl[z_{max}, \frac{\eta}{E(1-c)}\bigr] \ , \end{equation}
\begin{equation}\label{44} II_b  - region =  0 <
\theta-\arccos{\bigl(1-\frac{\eta}{zE}\bigl)}  <  \theta_2  <
\theta_m \ , \ \ \frac{\eta}{E(1-c)}  <  z  <  \bigl[z_{max},
\frac{\eta}{E(1-c_-)}\bigr] \ , \end{equation} where $c_- =
\cos{(\theta-\theta_m)}.$

Considering the integration limits  defined by the relations (39)--(44)
one can see that there is only one region $I_a$ in which the untagged
photon energy fraction can reach its minimal value $\Delta E.$ Because in
this region all angles for the untagged photon are allowed we can perform
an angular integration on the right--hand side of Eq.(37). The result
reads \begin{equation}\label{45} \frac{d\sigma^{^{H}}(I_a)}{dq^2} =
\frac{d\sigma^{^{B}}}{dq^2}\frac{\alpha}
{2\pi}\bigl[-(4\ln\Delta+3)L_m + 4\ln\Delta\bigr] +
\frac{d\sigma^{^{H}}_1}{dq^2} \ ,
\end{equation}
$$\frac{d\sigma^{^{H}}_1}{dq^2} =
\int\limits_0^{[z_{max},\frac{\eta}{E(1- c_+)}]}2
\frac{d\sigma^{^{B}}_{sh}}{dq^2}\frac{\alpha}{2\pi}[P_1(1-z, L_m)
-2A(1-z)]d\ z \ , \ L_m = \ln\frac{E^2\theta_m^2}{m^2},$$ where
$$P_1(x,L) = lim\ \Delta\rightarrow 0\ \
\bigl[\frac{1+x^2}{1-x}\theta(1-x-\Delta) + \bigl(\frac{3}{2}+2
\ln\Delta\bigr)\delta(1-x)\bigr]L \ , $$ $$A(x) = lim\
\Delta\rightarrow 0\ \ \frac{x}{1-x}\theta(1-x-\Delta) +
\ln\Delta\delta(1-x) \ .$$ In Eq.(45) we separate the dependence of the
contribution caused by the untagged hard photon emission on an auxiliary
parameter $\Delta.$ One can check explicitly that this term
together with (36) cancels the $\Delta$--dependence of the the soft and
virtual contribution(see Eqs.(27) and (30)).

We can also perform the analytical angular integration for the  contribution
of the region $I_b$ on the right--hand side of Eq.(37)
\begin{equation}\label{46, contribution of I_b}
\frac{d\sigma^{^{H}}(I_b)}{dq^2} =
\int\limits_{\frac{\eta}{E(1-c_+)}}^
{[z_{max},\frac{\eta}{E(1-c)}]}2\frac{d\sigma^{^{B}}_{sh}}{dq^2}\frac{\alpha}
{2\pi}\bigl[\frac{1+(1-z)^2}{z}\ln(1+\gamma) -
\frac{2(1-z)\gamma}{z(1+\gamma)} \bigr]d\ z \ ,
\end{equation}
$$ \gamma =
\frac{E^2}{m^2}\bigl[\arccos\bigl(1-\frac{\eta}{zE}\bigr)-
\theta\bigr]^2 \ .$$

Concerning the contribution of region $II$ in Eq.(37), we can
perform the corresponding analytical integration over the
azimuthal angle only. For the remaining variables $(z,\ \theta_2)$
we will show the limits of integration \begin{equation}\label{47,
contribution of region II} \frac{d\sigma^{^{H}}(II)}{dq^2} =
\Bigl\{\int\limits_{\frac{\eta}{E(1-c_+)}}^ {[z_{max},
\frac{\eta}{E(1-c)}]}dz\int\limits_{\arccos\bigl(1-\frac{\eta}
{zE}\bigr)-\theta}^{\theta_m}d\theta_2 +
\int\limits_{\frac{\eta}{E(1-c)}}^ {[z_{max},
\frac{\eta}{E(1-c_-)}]}dz\int\limits_{\theta - \arccos\bigl(1-
\frac{\eta}{zE}\bigr)}^{\theta_m}d\theta_2 \Bigr\}
\end{equation}
$$\times \frac{2}{\pi}\arccos{(B)}\frac{d\sigma^{^{B}}_{sh}}{dq^2}
\frac{\alpha}{2\pi}E^2\sin{\theta_2}\bigr[\frac{1+(1-z)^2}{(k_2p_1)}-
\frac{m^2z(1-z)}{(k_2p_1)^2}\bigr] \ . $$

Thus, the contribution in the RC due to the additional untagged hard photon
emission is given by the sum of Eqs. (36), (45), (46) and (47).

\vspace{0.5cm}

\section{Full first--order radiative correction}

\hspace{0.7cm}

The first--order radiative correction to the cross--section for
process (1) as measured by the KLOE detector with the above realistic
experimental restrictions can be written as
\begin{equation}\label{48}
\frac{d\sigma^{^{RC}}}{dq^2}= \frac{\alpha}{\pi}\sigma(q^2)\frac{s-q^2}
{4s^2}d\cos{\theta}\Theta\bigr(\frac{s-q^2}{4E}-\omega_m\bigr)
\frac{\alpha}{2\pi}\Bigl\{T +\frac{(s+t_1)^2+(s+t_2)^2}{t_1t_2}
\end{equation}
$$\times\Bigl[3\ln\frac{q^2}{s}+\frac{2\pi^2}{3}-\frac{9}{2}+(3+4\ln\Delta_1)
\ln\frac{4}{\theta_m^2} + 4\ln\frac{\Delta_2}{\Delta_1}\ln\frac{1+c_1}{1-c_1}
- \ln\Delta_2\ln\frac{1+c_2}{1-c_2}\Bigr]\Bigr\} +$$
$$\frac{d\sigma^{^{H}}_1}{dq^2} + \frac{d\sigma^{^{H}}(I_b)}{dq^2} +
\frac{d\sigma^{^{H}}(II)}{dq^2} \ , $$
where we used the approximation $\cos{\theta_m} = 1 -\theta_m^2/2.$

The total cross--section $\sigma(q^2)$ of the one--photon
annihilation process $e^+e^-\rightarrow \pi^+\pi^-$ which should
be extracted from the KLOE experiment measurements, is factorized on
the right side of Eq.(48). This follows from the expression for
$d\sigma^B_{sh}/dq^2$, see Eq.(38), which enters into each term in
the third line in Eq.(48). This demonstrates an evident advantage
of the approach of the Ref. [6] as compared with the scanning of
the hadron cross--section by the tagged photon energy
measurements. In the latter approach the RC caused by the
additional invisible hard photon radiation include with necessity
some integrals over $\sigma(q^2)$ [15,16]. These integrals arise
because in this case the tagged photon energy does not define the
pion invariant mass directly.

As noted above, the auxiliary parameter $\Delta$ disappears
from the master formula (48) for the radiative correction to the
cross--section of process (1). But all physical parameters,
which define the  event selection (namely: the "softness"
parameters $\Delta_1$ and $\Delta_2,$ angular parameters
$\theta_m, \ \theta_1$ and $\theta_2$ as well as the energy threshold
$\omega_m$ and parameter $\eta$) enter this formula either
explicitly or via the integration limits in the third line. The
differential cross--section over the measured $\pi^+\pi^-$--invariant mass
$q^2$ is given by the sum of the Born term (17) which
depends on the parameter $|\vec P_{\Phi}|$, and (48)
\begin{equation}\label{49}
\frac{d\sigma}{dq^2} =
\frac{d\sigma^{^{B}}}{dq^2} + \frac{d\sigma^{^{RC}}}{dq^2} \ .
\end{equation}

\section{Conclusion}

\hspace{0.7cm}
A crucial requirement for success of the forthcoming precision studies of
the hadronic cross--section $\sigma(e^+\,e^-\rightarrow hadrons)$
at DA$\Phi$NE through the measurements of
radiative events [6,7] is the matching level of
reliability of the theoretical expectation. This, in turn, requires a
detailed knowledge of the radiative corrections corresponding to the
realistic conditions of the KLOE detector.

In this paper we derive the analytical expressions for
the distribution over the invariant mass of the charged pion pair
corresponding to the constraints of the proposed experiment [6] with the
KLOE detector. When DA$\Phi$NE operates at the $\Phi$ peak just this
two--pion final hadronic state provides the dominant contribution to the
ISR events due to the radiative return to the $\rho$ resonance. Our approach
can be quite straightforwardly extended to the description of ISR
events in the general case of an arbitrary  hadronic final state.

Our formulae take into account both the kinematical constraints related
to the geometry of the photon detector and the event selection cuts
imposed in order to reduce the FSR contamination. When deriving the Born
results the Lorentz boost of the $\Phi$ in the laboratory frame was
accounted for.  First of all, such an accuracy is essential for the
high--precision determination of the tagged photon energy. For the
purposes of calculation of the RC with the one per cent accuracy the
$|\vec P_{\Phi}|/2E$ effects may be neglected.

A prospective advantage of the experimental strategy proposed in [6] is
the direct precise determination of the two--pion invariant mass which, in
turn allows to avoid the deconvolution procedure. One may even think about
taking full advantage of the high precision of the measurements of the two
charged pions with the drift chamber[6] by making the photon tagging
redundant \footnote{We are grateful to G. Venanzoni who has attracted our
attention to such an option.}.  An obvious attractiveness of such an
inclusive strategy is that the "invisible" ISR photons are then emitted
 dominantly in the very forward
cones along the beams and the corresponding cross--sections are large
(due to $\ln{E^2/m^2}$ enhancement). The overall event geometry becomes
rather simple and the corresponding RC are governed by quasireal
kinematics, see [21].
Then the constraints imposed by the performance
of KLOE calorimeters become unimportant.

The corresponding results for the Born cross--section were presented in
Ref. [14]. The derivation of the RC requires some modifications (as
compared to the results given in Ref. [14]) due to the contribution from
additional hard photon radiation, since, in this case, the invariant mass
of the pions and not the energy of the tagged photon is measured.

However, the success of such an
inclusive approach requires a special care regarding different background
events. Thus, a carefully chosen event selection should be introduced in
order to reduce as much as possible various contaminations such as FSR
events, $\Phi \rightarrow \pi^+\,\pi^-\,\pi^0 \ ; \pi^+\,\pi^-\,\gamma$
etc as well as double--photon mechanism of $\pi^+\,\pi^-$ production.
Further detailed
examination of the background caused by the strong decay modes and
especially by the contribution of double--photon $\pi^+\,\pi^-$ production
has to be performed. We plan to perform these studies elsewhere. \\

\vspace {0.5cm}

{\large{\bf Acknowledgements}} \\

\hspace{0.7cm} Authors thanks V.S. Fadin, G. Venanzoni and W.
Kluge for fruitful discussion and critical remarks. N.P.M. thanks
INFN for the hospitality. V.A.K. thanks the Leverhulme Trust for a
Fellowship. This work was partly supported by the EU Framework TMR
programme, contract FMRX--CT98--0194 (DG 12--MIHT) and CT--98--0169.

\vspace{0.8cm}

{\large{\bf References}}
\begin{enumerate}
\item S. Aid et al., H1 Coll., Nucl.Phys. {\bf{B 470}} (1996) 3.
\item M.W. Krasny, W. Plazcek, H. Spiesberger, Z.Phys. {\bf{C 53}} (1992)
687.
\item D. Bardin, L. Kalinovskaya, T. Riemann, Z.Phys. {\bf{C 76}} (1997)
487; \\ H. Anlauf, A.B. Arbuzov. E.A. Kuraev, N.P. Merenkov, JETP
Lett. {\bf 66} (1997) 391, {\it ibid} {\bf 67} (1998) 305; JHEP
9810 (1998) 013; Phys.Rev. {\bf D 59} (1999) 014003. \\ H. Anlauf,
A.B. Arbuzov, E.A. Kuraev. In "Hamburg 1998/1999, Monte Carlo
generators for HERA physics", p.539, hep--ph 9907248; \\ H.  Anlauf, Eur.
Phys.  J.  {\bf C 9} (1999) 69.  \item M.  Benayoun, S.I.  Eidelman, V.N.
Ivanchenko, Z.K.  Silagadze, hep--ph/9910523.  \item For recent
publication see, for example, L3 Collab., M. Acciarri, et al., Preprint
CERN--EP/99--129, Sept.  1999.  \item G. Cataldi, A. Denig, W.  Kluge, G.
Venanzoni, KLOE MEMO 195, August 13, 1999.  \item S.  Spagnolo, "{\it La
camera a drift di KLOE e prospettive e obiettivi di una nuova misura delle
correzioni adroniche al g-2 muone a DA$\Phi$NE"}, Doctorate thesis (in
Italian), University of Lecce, 1999, unpublished; \\ S.  Spagnolo, Eur.
Phys. J. {\bf C 6} (1999) 637.  \item S. Eidelman, F.  Jegerlehner, Z.
Phys. {\bf C 67} (1995) 585; \\ F. Jegerlehner, {\it "Precision
measurement of hadronic cross-sections at low energies"}, \\
EURODA$\Phi$NE Meeting, LNF, Frascati, Italy (1998); \\ F. Jegerlehner,
{\it "Hadronic effects in $(g-2)_{\mu}$ and $\alpha_{QED}(M_Z)$: Status
and perspectives}, Preprint DESY 99-007, hep-ph/9901386; \\ F.
Jegerlehner, {\it "Sigma Hadronic and Precision Tests of the SM"}, LNF
Spring School and EURODA$\Phi$NE Collaboration Meeting, Frascati, Italy,
(1999).  \item N.  Cabibbo, R.  Gatto, Phys.  Rev.  {\bf 124} (1961) 1577.
\item CDM--2 Collaboration, R.R. Akhmetshin et al., Preprint
Budker INP--99--10, \\ hep--ex/9904027.
\item V.N.  Baier, V.A. Khoze, Sov. Phys.  JETP {\bf 21} (1965) 629, 1145.
\item  G.~Pancheri, Nuovo Cim. A {\bf 60}, 321 (1969);\\
M.~Greco, G.~Pancheri and Y.N.~Srivastava, Nucl. Phys. {\bf B101}, 234 (1975).
\item M.J. Greutz, M.B. Einhorn,
Phys. Rev. {\bf D 9} (1970) 2537;
\\ G. Bonneau, F. Martin, Nucl. Phys. {\bf B 27} (1971) 381; \\ M.  Greco,
Riv. Nuovo Cim. {\bf 11} (1988) 1.  \item A.B. Arbuzov, E.A. Kuraev, N.P.
Merenkov, L. Trentadue, JHEP 12 (1998) 009.  \item M. Konchatnij, N.P.
Merenkov, JETP Lett. {\bf 69} (1999) 811.  \item S. Binner, J.H.
K\H{u}hn and K. Melnikov, Phys. Lett. {\bf B 459} (1999) 279.  \item
M. Caffo, H. Czyz and E. Remiddi, Nuovo Cim. {\bf 110 A} (1997) 515;
Phys. Lett. {\bf B 327} (1994) 369.  \item E.A. Kuraev, N.P. Merenkov,
V.S. Fadin, Sov. J. Nucl. Phys. {\bf 45} (1987) 486.  \item A.B.
Arbuzov et al., Nucl. Phys. {\bf B 485} (1997) 457.  \item E.A. Kuraev
and V.S.Fadin, Sov. J. Nucl. Phys. {\bf 41} (1985) 466.  \item V.N.
Baier, V.S. Fadin, V.A. Khoze, Nucl. Phys. {\bf 65} (1973) 381.

\end{enumerate}

\end{document}